\def\year{2015}
\documentclass[letterpaper]{article}
\usepackage{aaai}
\usepackage{times}
\usepackage{helvet}
\usepackage{courier}
\usepackage{amssymb}
\usepackage{amsmath}
\usepackage{algorithm}
\usepackage{algpseudocode}
\usepackage{graphicx}
\usepackage{float}

\begin{document}
%
\title{Exploiting Consistency Theory for Modeling Twitter Hashtag Adoption}
\author{Hamidreza Alvari\\
Department of EECS, University of Central Florida, Orlando, FL, USA\\
halvari@eecs.ucf.edu
}
\maketitle
\begin{abstract}
\begin{quote}
Twitter, a microblogging service, has evolved into a powerful communication platform with millions of active users who generate immense volume of microposts on a daily basis. To facilitate effective categorization and easy search, users adopt hashtags, keywords or phrases preceded by hash (\#) character. Successful prediction of the spread and propagation of information in the form of trending topics or hashtags in Twitter, could help real time identification of new trends and thus improve marketing efforts. Social theories such as consistency theory suggest that people prefer harmony or consistency in their thoughts. In Twitter, for example, users are more likely to adopt the same trending hashtag multiple times before it eventually dies. In this paper, we propose a low-rank weighted matrix factorization approach to model trending hashtag adoption in Twitter based on consistency theory. In particular, we first cast the problem of modeling trending hashtag adoption into an optimization problem, then integrate consistency theory into it as a regularization term and finally leverage widely used matrix factorization to solve the optimization. Empirical experiments demonstrate that our method outperforms other baselines in predicting whether a specific trending hashtag will be used by users in future.  
\end{quote}
\end{abstract}

\noindent Twitter\footnote{https://twitter.com}, a prevalent and well-known microblogging Website allows millions of active users interacting with each other and posting tweets, a message up to 140 characters, per day on computers or mobile devices. Twitter is popular for massive spreading of tweets and the nature of freedom. Daily bursts of news, gossips, rumors, discussions and many others are all exchanged and shared by users all over the world, no matter where they come from, civilized or uneducated, or even what religion they hold. Consequently, users on Twitter are easily overwhelmed by the tremendous volume of data. 

To ease the task of categorization and following up with the trends, Twitter has allowed users to freely assign valid hashtags to their tweets, i.e. strings prefixed by the hash "\#" character. Hahstags could help users categorize their own posts and thus represent a coarse-grained topic of the content. This mechanism is a community-driven convention for adding additional context to tweets. In particular, it helps tweet search and quickly propagation of the topic among millions of users by allowing them to join the discussion. Hashtags could be viewed as topical markers to indicate the core idea expressed in the tweet and hence are adopted by users who contribute similar content. Trending hashtags are those hashtags which receive extensive attention in a strictly short period of time due to certain reasons but eventually die at some point. Understanding the spread and propagation of such information through Twitter has many immediate applications such as targeting users for marketing purposes, identification of trends and enhancing marketing efforts, or even socio-political events and large natural disasters. 

Consistency theory~\cite{Abelson1983} is a social theory which suggests that people prefer harmony or consistency in their inner systems (beliefs, attidues, thoughts, etc.). In other words, when things fall out of alignment, the discomfort of cognitive dissonance occurs to help people keep their practical level of consistency in their lives by motivating them to change their thoughts to restore consistency. An example of the consistency theory in Twitter is that the hashtags adopted by the same user are more likely to be consistent than those of two randomly chosen hashtags. Based on the consistency theory, we envision that users that have previously adopted a certain trending hashtag in past, are more likely to adopt it again in future. 

In this study we aim at modeling the information spread in the form of trending hashtag adoption in Twitter based on matrix factorization scheme and consistency theory. Our main contributions are then as follows: 
\begin{itemize}
	
	\item We perform two-sample \textit{t}-test to verify that users are more likely to adopt the same hashtag multiple times and hence possess consistent hashtag usage history.
	
	\item We formulate the problem of trending hashtag adoption prediction into an optimization problem and integrate consistency theory into it. To take into account the fact that trending hashtags do not last long, we further incorporate {\it attenuation} matrix into the optimization equation. We use low-rank weighted matrix factorization model to solve the optimization equation and propose $hCWMF$. To accommodate the process of optimization and fast finding of suboptimal matrices, we use alternating least square scheme for updating the corresponding matrices.
	 
	\item We collect and build a dataset of tweets of 6 different trending hashtags to evaluate the proposed model and demonstrate its ability to predict the hashtag adoption by users.
\end{itemize}

This paper is organized as follows. In the first section, we explain our data crawling methodology.  Next, we provide formal definition of the problem in hand and notations used throughout the paper and detail our matrix factorization framework and its time complexity. We conduct the experiments and discuss the results in the next section and finally conclude the paper with conclusion and future work section.

\section{Data}
We collect and build our dataset by using the Twitter streaming API\footnote{https://dev.twitter.com/streaming/overview} which provides 1\% random tweets from the total volume of tweets at a particular moment. In general, collecting the appropriate and standard dataset requires heavy effort along with human annotation to evaluate and verify the data which obviously is a tedious task. This makes evaluating the hashtag adoption prediction approach even harder. Thus, we need a more automatic and systematic way of generating the dataset. Furthermore, we rather preferred topics that have been trending only for a strictly short period of time (less than a week) in order to support having several spikes. Therefore we collected tweets from 6 different trending topics including: CopenHagenShooting (12,994 tweets), MesaShooting (2,659 tweets), ChapelHillShooting (136,067 tweets), JeSuisCharlie (15,620 tweets), OscarsSoWhite (9,430 tweets), GoodAdviceIn4Words (248,961 tweets) each of which were trending for at most a week between Jan 2015 and March 2015. 

\begin{figure}[t]
	\centering
	\includegraphics[width=0.5\textwidth]{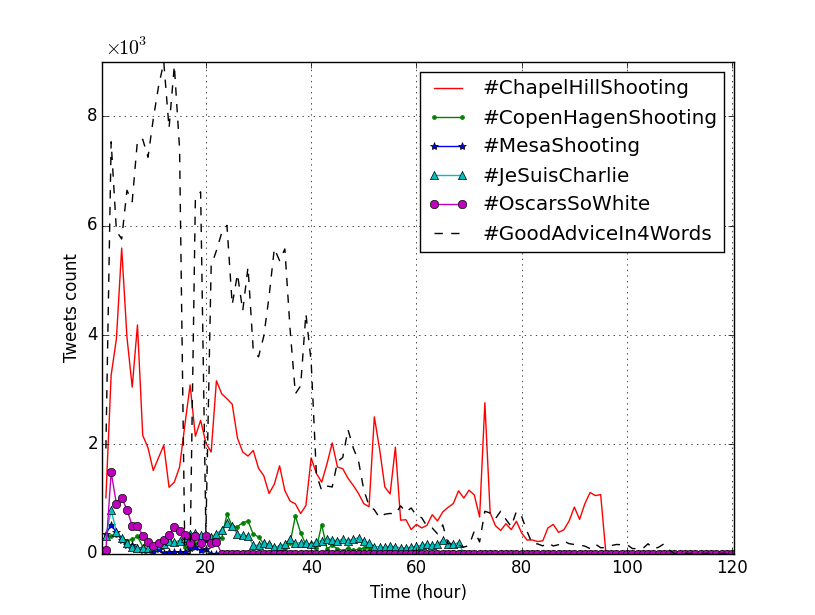}
	\caption{Tweet counts of each trending topic. Each hashtag has different number of spikes for the same amount of time.}\label{fig:alltweets}
\end{figure}

\begin{figure}[H]
	\centering
	\includegraphics[width=0.5\textwidth]{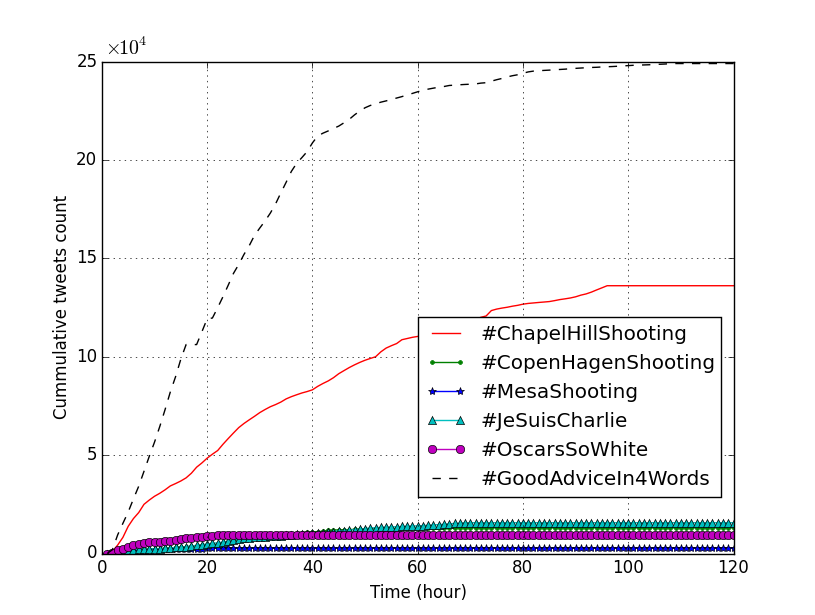}\\
	\includegraphics[width=0.5\textwidth]{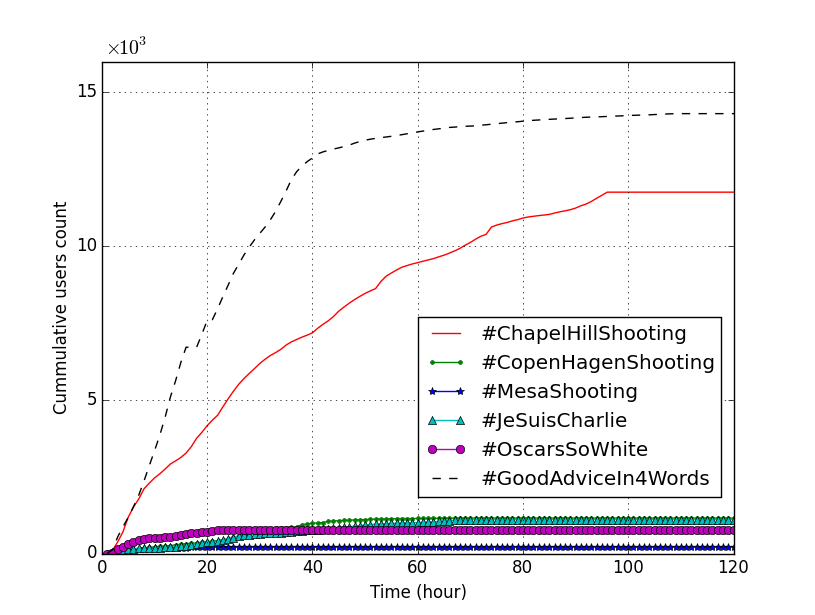}
	\caption{Cumulative number of tweets and users over time}\label{fig:cdfs}
\end{figure}

The description of the resulting dataset which has about \%0.99 sparsity is shown in Table~\ref{tb:data}. Also, few anonymized tweet examples of the dataset are shown in Table~\ref{tb:tweets}. We further demonstrate hourly tweet counts of each trending topic in Fig.~\ref{fig:alltweets}. As expected, users behave differently while posting in different topics in the same period of time. In more details, number of tweets and thus user participating in each topic and also the pace at which users write on different topics vary greatly. Another observation is that none of the trending topics lasts long and they all eventually die at around the fifth day of data collection. We also demonstrate cumulative number of tweets and users over time in Fig.~\ref{fig:cdfs} for all hashtags in our dataset in order to show the user participations in different topics of our dataset. 

\begin{table}[t]
	\centering
	\caption{Description of dataset}
	\label{tb:data}
	\begin{tabular}{l c}\hline\hline
		\textbf{\# of Trending Hashtags} & 6 \\
		\textbf{\# of Users} &  212,062\\
		\textbf{\# of Tweets} & 425,731 \\
	\end{tabular}
\end{table}

\begin{table*}[t]
	\centering
	\caption{Examples of collected tweets}
	\label{tb:tweets}
	\begin{tabular}{l p{12cm}}\hline
		\textbf{Hahstag} & \textbf{Tweets} \\\hline
		\textbf{\#ChapelHillShooting} & RT @...: Thousands lay \#ChapelHillShooting victims to rest http://... \#MuslimLivesMatter http://... \\
		\textbf{\#OscarsSoWhite} &  The Oscars' lack of diversity is depressing. And no one's doing anything to change it. http://... \#OscarsSoWhite \\
		\textbf{\#GoodAdviceIn4Words} & Love more, hate less. \#GoodAdviceIn4Words \\
	\end{tabular}
\end{table*}

\section{Problem Formulation and Notations}
Before going further, we provide the formal definition of the problem and the notations used in the paper. 

Given a hashtag $H$ and a set of $N$ users who have adopted $H$ in past during time intervals $T=\{[t_1^i,t_2^i] \mid \forall i = 1,...,N \}$, we aim at predicting the next times $T' = \{t'^i \mid t'^i > t_2^i, \forall i=1,...,N\}$ in future that each user will adopt $H$ when posting tweets on Twitter. Suppose we have a very sparse and low-rank user-time matrix $\textbf{X}=[x_{ij}] \in \mathbb{R}_{+}^{N\times M}$ where $M \gg max \{t_2^i \mid \ i = 1,...,N\}$. We denote by $x_{ij}, 1\leq i \leq N, 1\leq j \leq M$ the $i$th row and $j$th column of $\textbf{X}$, which represent if user $i$ has adopted hashtag $H$ at time $j$. The reason that $\textbf{X}$ is sparse is because in Twitter, users do not usually adopt hashtags while tweeting and despite the availability of this feature, only 8\% of the tweets contain hash "\#" character \cite{kywe2012recommending}. 

We formulate the problem of hashtag adoption prediction into an optimization problem and employ low-rank matrix factorization to solve it as this method has been widely and successfully employed in various applications such as collective filtering \cite{koren2008factorization} and document clustering \cite{Zhu07}. In its basic form and in the context of recommendar systems, matrix factorization, one of the realizations of latent factor models, captures both items and users by vectors of factors inferred from the ratings. In this study we characterize users and hashtag adoption times by inferring vectors of factors from user's hashtag usage history. 

\subsection{Matrix factorization model}
Based on matrix factorization scheme, we seek two low-rank and non-negative matrices $\textbf{U} \in \mathbb{R}_{+}^{N\times d}$ and $\textbf{V} \in \mathbb{R}_{+}^{M\times d}$ with dimensionality of the latent space $d \ll M,N$ via solving the following optimization problem:

\begin{equation}
\begin{split}
\label{eq:main_opt}
 min_{\textbf{U},\textbf{V}} ||\textbf{W} \odot (\textbf{X} - \textbf{UV}^T)||^2_F + \gamma_1||\textbf{U}||^2_F + \gamma_2||\textbf{V}||^2_F \\ + \mu||\textbf{G} \odot (1-\textbf{UV}^T)||^2_F
\end{split}
\end{equation}

where $\odot$ is Hadamard product (element-wise product) where $(\textbf{X}\odot \textbf{Y})_{ij} = \textbf{X}_{ij} \times \textbf{Y}_{ij}$ for any two matrices $\textbf{X}$ and $\textbf{Y}$ with the same size, $||\textbf{.}||_F$ is the Frobenius norm of a matrix, $||\textbf{A}||_F=\sqrt{\sum_i \sum_j \textbf{A}_{ij}^2}$, $\textbf{W}=[w_{ij}] \in \mathbb{R}_{+}^{N\times M}, 1\leq i \leq N, 1\leq j \leq M$ is an indicator matrix to control the learning process and finally $\textbf{G}=[g_{ij}] \in \mathbb{R}_{+}^{N\times M}, 1\leq i \leq N, 1\leq j \leq M$ is an attenuation matrix to take into account the fact that trending hashtags do not last long. Also, $\gamma_1 >0 $ and $\gamma_2 > 0$ are non-negative regularization parameters and $||\textbf{U}||^2_F$ and $||\textbf{V}||^2_F$ are two smoothness regularization terms to avoid overfitting. The row vectors $u_{i.}, 1\leq i \leq N$ and  $v_{j.}, 1\leq j \leq M$ denote the low-dimensional representations of users and adoption times respectively.

We integrate $\textbf{W}$ into the optimization equation to avoid impacts of unknown elements of $\textbf{X}$, i.e. increase the contribution of the elements with known values in the optimization process over the elements with the missing information. Therefore, for those times that we exactly know that a user has adopted hashtag $H$ or not (training set), we use $w_{ij}=1$ and for those with missing information (test set) we set $w_{ij}=0$. In other words, the indicator matrix is formally defined as: 

\begin{equation}\label{eq:w}
w_{ij}=\begin{cases}
1, & x_{ij} \in \textbf{X}\\
0, & otherwise
\end{cases}
\end{equation}

We further define attenuation matrix $\textbf{G}$ in the following way. In each row $i$ of $\textbf{X}$, we proceed to find the first entry ($j$-th column) which equals to 1, i.e. $x_{ij}=1$. Once found, we set the corresponding entry in $\textbf{G}$ to 1 and define others as follows:

\begin{equation}\label{eq:g}
g_{ij}=\begin{cases}
0, & \forall k < j\\
1, & x_{ij} = 1\\
1 - \frac{1}{M - j + 1}, & \forall k > j
\end{cases}
\end{equation}

In other words, the remaining entries of \textbf{G} that come after the first seen element equals to one, would be all less than one and in a decreasing order, indicating the attenuation of time as it goes by.  

The latter part of Eq.~\ref{eq:main_opt} refers to the modeling of the consistency theory. According to this social theory, Twitter users prefer consistency in their attitudes and thoughts and hence they are more likely to adopt the same hashtag over and over. Furthermore, hashtags do not last long, therefore we incorporate $\textbf{G}$ with the larger values indicating the earlier adoption of a certain hashtag. As time goes by, the probability of adopting the same hashtag attenuate, but at the same time some users could be still interested to adopt it in future based on the consistency theory which forces $\textbf{UV}^T$ to approach to 1.

\textbf{Optimization.} The coupling between $\textbf{U}$ and $\textbf{V}$ in the optimization problem, makes it difficult to find the optimal solutions for both matrices. Therefore, in this work, we adopt the alternating least squares method\cite{conf/icdm/DingLJ08} to solve the optimization problem, where the objective function is iteratively optimized with respect to one of the variables $\textbf{U}$ and $\textbf{V}$ while fixing the other one until convergence. Optimizing Eq.~\ref{eq:main_opt} with respect to $\textbf{U}$ and $\textbf{V}$ corresponds to the computation of their derivatives via the following equations. Given the following objective function:

\begin{equation}
\begin{split}
\label{eq:final_opt}
\textbf{L} = ||\textbf{W} \odot (\textbf{X} - \textbf{UV}^T)||^2_F + \gamma_1||\textbf{U}||^2_F + \gamma_2||\textbf{V}||^2_F \\ + \mu||\textbf{G} \odot (1-\textbf{UV}^T)||^2_F
\end{split}
\end{equation}

the update equations for $U$ and $V$ are computed according to the following equations:
\begin{equation}\label{eq:UUpdate}
\textbf{U} = \textbf{U} - \lambda \frac{\partial \textbf{L}}{\partial \textbf{U}}
\end{equation}
\begin{equation}\label{eq:VUpdate}
\textbf{V} = \textbf{V} - \lambda \frac{\partial \textbf{L}}{\partial \textbf{V}}
\end{equation}
where $\lambda > 0$ is the learning step and the partial derivatives of $\textbf{L}$ with respect to $\textbf{U}$ and $\textbf{V}$ are then obtained using:

\begin{equation}\label{eq:partialU}
\begin{split}
\frac{\partial \textbf{L}}{\partial \textbf{U}} = -2\textbf{(W}\odot \textbf{X)V} + 2\textbf{(W}\odot(\textbf{UV})^T)\textbf{V} + \gamma_1\textbf{U} \\ - \mu[ 2\textbf{G}\textbf{V} + 2(\textbf{G} \odot (\textbf{UV}^T))\textbf{V}]
\end{split}
\end{equation}

\begin{equation}\label{eq:partialV}
\begin{split}
\frac{\partial \textbf{L}}{\partial \textbf{V}} = -2(\textbf{W}\odot \textbf{X})^T\textbf{U} + 2(\textbf{W}\odot(\textbf{UV}^T))^T\textbf{U} + \gamma_2\textbf{V} \\ - \mu[2\textbf{G}^T\textbf{U} + 2(\textbf{G} \odot (\textbf{UV}^T)^T)\textbf{U}]
\end{split}
\end{equation}

Upon the convergence, we approximate $\textbf{X}$ by multiplying the low-rank matrices $\textbf{U}$ and $\textbf{V}$:
\begin{equation}
\widetilde{\textbf{X}} \approx \textbf{UV}^T
\end{equation}

\textbf{Algorithm.} The detailed algorithm for the proposed matrix factorization framework is shown in Algorithm~\ref{alg:alg1}. After randomly initializing matrices $\textbf{U}$ and $\textbf{V}$ and constructing the indicator matrix $\textbf{W}$ and attenuation matrix $\textbf{G}$ in lines 3-6, in lines 8-11, we alternatively update $\textbf{U}$ and $\textbf{V}$ based on the equations~\ref{eq:UUpdate} to~\ref{eq:partialV}, until we reach convergence. Practically, convergence is achieved whenever predefined maximum number of iterations has been reached or there is little change in the objective function value. Finally,  $\widetilde{\textbf{X}} \approx \textbf{UV}^T$ is the low-rank representation of user-time matrix $\textbf{X}$ and also is non-negative as $\textbf{U}$ and $\textbf{V}$ are both non-negative matrices. 

\begin{algorithm}
	\caption{The proposed framework of hashtag adoption prediction}\label{alg:alg1}
	\begin{algorithmic}[1]
		\State \textbf{Input:} User-time matrix $\textbf{X}$, $d$, $\gamma_1$, $\gamma_2$, $\lambda$
		\State \textbf{Output:} Modeled matrix $\widetilde{\textbf{X}}$
		\State \text{Initialize \textbf{U} randomly}
		\State \text{Initialize \textbf{V} randomly}
		\State \text{Construct the indicator matrix $\textbf{W}$ according to eq.~\ref{eq:w}}
		\State \text{Construct the matrix \textbf{G}} according to eq.~\ref{eq:g}
		\While{\text{Not convergent}}
		\State $\frac{\partial \textbf{L}}{\partial \textbf{U}} = -2\textbf{(W}\odot \textbf{X)V} + 2\textbf{(W}\odot(\textbf{UV})^T)\textbf{V} + \gamma_1\textbf{U} - \mu[ 2\textbf{G}\textbf{V} + 2(\textbf{G} \odot (\textbf{UV}^T))\textbf{V}]$
		\State \text{Update} $\textbf{U}\gets \textbf{U} - \lambda  \frac{\partial \textbf{L}}{\partial \textbf{U}}$
		\State $\frac{\partial \textbf{L}}{\partial \textbf{V}} = -2(\textbf{W}\odot \textbf{X})^T\textbf{U} + 2(\textbf{W}\odot(\textbf{UV}^T))^T\textbf{U} + \gamma_2\textbf{V} - \mu[2\textbf{G}^T\textbf{U} + 2(\textbf{G} \odot (\textbf{UV}^T)^T)\textbf{U}]$
		\State \text{Update} $\textbf{V}\gets \textbf{V} - \lambda  \frac{\partial \textbf{L}}{\partial \textbf{V}}$
		\EndWhile
		\State \text{Set $\widetilde{\textbf{X}} = \textbf{UV}^T$}
	\end{algorithmic}
\end{algorithm}

\subsection{Time Complexity}
We discuss the time complexity of the proposed method here. Obviously, the complexity burden of our method depends mostly on the computation of the derivatives in Equations~\ref{eq:partialU} and ~\ref{eq:partialV}. In each iteration in algorithm~\ref{alg:alg1}, the time complexities of the computation of the derivatives in lines 8 and 10 are calculated as follows: first note that $\textbf{W}\odot \textbf{X}$, $\textbf{UV}^T$ and $\textbf{(W}\odot(\textbf{UV}^T))$ need to be calculated once for both equations. The time complexity of $\textbf{W}\odot \textbf{X}$ is $\mathcal{O}(N_x)$ where $N_x$ is the number of non-zero elements of the sparse matrix $\textbf{X}$. Also, the time complexity of $\textbf{UV}^T$ and $\textbf{GV}$ are both $\mathcal{O}(NdM)$. For $\textbf{(W}\odot \textbf{X)V}$, we need $\mathcal{O}(N_xd)$. For the second and last terms in Eq.~\ref{eq:partialU}, i.e. $\textbf{(W}\odot(\textbf{UV}^T))\textbf{V}$, $\textbf{(G}\odot(\textbf{UV}^T))\textbf{V}$, we need $\mathcal{O}(NM + NMd)$. Thus in each iteration, the calculation of $\frac{\partial \textbf{L}}{\partial \textbf{U}}$ takes $\mathcal{O}(N_xd + NMd)$. With the similar computations, the calculation of $\frac{\partial \textbf{L}}{\partial \textbf{V}}$ has the time complexity of $\mathcal{O}(N_xd + NMd)$ in each iteration.

\subsection{\textit{t}-test}
We perform two-sample \textit{t}-test to verify the existence of consistency theory in our data and answer the question: \textit{Do users in Twitter use the same hashtag multiple times?} 

We construct two vectors $hc_u$ and $hc_r$ with the equal number of elements where each element in $hc_u$ is obtained by simply counting hashtags that have been used by user $u$ multiple times and similarly each element in $hc_r$ is the number of the same hashtags used by user $u$ and a random user $r$. 

We perform a \textit{t}-test on vectors $hc_u$ and $hc_r$. The null hypothesis here is that the number of same hashtags used by a given user is less than or equal to those of different users, i.e. $H_0: hc_u \leq hc_r$, while the other hypothesis is that the number of same hashtags used by the same user is more than those of different users, $H_1: hc_u > hc_r$. Therefore we have the following two-sample \textit{t}-test:
\begin{equation}
	H_0: hc_u \leq hc_r, H_1: hc_u > hc_r
\end{equation}
The \textit{t}-test result suggests a strong evidence with the significance level $\alpha = 0.01$ with p-value 2.53e-49 to reject the null hypothesis and as a result, confirms that users tend to use a the same hashtag multiple times. Therefore the answer to the above question is positive. This aligns well with our findings and equations in our matrix factorization model in the previous section.

\section{Experiments}
We conduct experiments to compare the performance of our proposed method with the baselines. In this section, we begin by introducing the evaluation metric we have used and then we design the experiments and discuss the results.

\subsection{Evaluation Metric}
We use widely used metric for evaluating collaborative filtering results, root mean square error (RMSE), which is defined as:
\begin{equation}
RMSE = \sqrt{\frac{\sum_{i,j}^n(\widetilde{x}_{ij} - x_{ij})^2}{n}}
\end{equation}

where $n$ is the number of test instances used for the evaluation and $\widetilde{x}_{ij}$ and $x_{ij}$ are corresponding test set elements of $\widetilde{X}$ and $X$ selected for the evaluation respectively.
\subsection{General Discussion}
To the best of our knowledge, there exists no similar method in the literature for predicting the hashtag adoption, therefore, we build the following baselines and compare the results as follows,

\textbf{WMF}: A variant of the proposed method without regularization term corresponding to the consistency theory.

\textbf{Autoregressive Moving Average (ARMA)}: this is a widely used time series analysis technique. We use statsmodels\footnote{http://statsmodels.sourceforge.net/} python package to conduct experiments using ARMA method. Given a time series of data $X_t$, the ARMA model seeks to understand and predict the future values in this series. ARMA has two parts, an autoregressive (AR) and a moving-average (MA) parts and is of the general form:
\begin{equation}
X_t = c + \epsilon_t + \sum_{i=1}^p\phi_iX_{t-i}+\sum_{i=1}^q\theta_i\epsilon_{t-i}
\end{equation}
where $\phi_1,...\phi_p$ and $\theta_1,...\theta_q$ are the parameters of AR and Moving-average models respectively, $c$ is a constant and $\epsilon_t$ is white noise.  

\textbf{Markov Chain (MC)}: first order markov chain is a sequence of random variables $X_1,...,X_n$ with Markov property, i.e. the current state only depends on the immediate past state. Formally,
\begin{equation}
\begin{split}
Pr(X_{n+1}=x|X_1=x_1,...,X_n=x_n)= \\ Pr(X_{n+1}=x|X_n=x_n)
\end{split}
\end{equation}
In more details, we consider a state diagram with two states 0 and 1 to indicate the probabilities of transitioning from 0 to 1 or vice versa in the user-time matrix \textbf{X}. We seek to estimate the probability of transitioning from 0 to 1 to indicate the probability that user adopts hashtag $H$ at time $t$. However, since the matrix \textbf{X} is strictly sparse, the probability is much higher for transitioning from either 0 or 1 to 0, in the transition matrix, thereby increasing the chance of getting stuck in the state 0 and failing to predict the correct values for unseen data. An example of a transition matrix for \#JeSuisCharlie is as follows,

\begin{equation}\nonumber
\textbf{T}=
\begin{bmatrix}
0.7813 & 0.0148 \\
0.8702 & 0.0122
\end{bmatrix}
\end{equation}

\textbf{Random}: this method randomly assigns binary values to the test instances as their predicted values independently from the data.

For our proposed method, we try various parameters and report the best performance, while other methods do not have parameters. In particular, we set both regularization parameters $\gamma_1$ and $\gamma_2$ to 0.2 and set the learning step $\lambda$ to 0.001. Also, we apply different dimensions of latent space $d$ and observe the best performance is achieved when $d=10$. For brevity, we only demonstrate this in Table~\ref{tb:results2} for \#MesaShooting data, when percentage of test set is fixed to \%30.

With the parameters chosen as above, our experimental setting is as follows: Suppose we have $\ell = \{ (u_i,t_j) \mid x_{ij} = 1 \}$ is the set of pairs of users and times that we know they have adopted particular hashtag $H$ at those time. We choose \%$x$ of $\ell$ as new relations $\mho$ between users and times to predict. We remove these relations by setting $x_{ij} = 0, \forall (u_i,h_j) \in \mho$ and then apply the hashtag adoption prediction approaches on the new representation of \textbf{X}. We vary $x$ as $\{10,20,30,40,50\}$. 

\begin{table*}[t]
	\centering
	\caption{Performance Comparison for different approaches in terms of RMSE with d = 10 and different testbed sizes}
	\label{tb:results1}
	\begin{tabular}{l|c|c|c|c|c|c}\hline
		Dataset & Method & \%10 & \%20 & \%30 & \%40 & \%50 \\\hline\hline
		\#CopenHagenShooting & hCWMF & \textbf{0.0798} & \textbf{0.0851} & \textbf{0.0891} & \textbf{0.0962} & \textbf{0.0983} \\
		& WMF & 0.2467 & 0.2574 & 0.2741 & 0.3045 & 0.3151  \\
		& ARMA & 0.9812 & 0.9852 & 0.9913 & 0.9998 & 1.0\\
		& MC & 1.0 & 1.0 & 1.0 & 1.0 & 1.0\\
		& Random & 0.7018 & 0.7076 & 0.7135 & 0.7047 & 0.7109  \\\hline
		\#MesaShooting & hCWMF & \textbf{0.1016} & \textbf{0.1160} & \textbf{0.1181} &  \textbf{0.1257} & \textbf{0.1301} \\
		& WMF & 0.2352 & 0.2704 & 0.2760 & 0.3073 & 0.3158 \\
		& ARMA & 0.9634 & 0.9787 & 0.9913 & 1.0 & 1.0\\
		& MC & 1.0 & 1.0 & 1.0 & 1.0 & 1.0\\
		& Random & 0.7200 & 0.7067 & 0.7054 & 0.7181 & 0.6078 \\\hline
		\#ChapelHillShooting & hCWMF & \textbf{0.0676} & \textbf{0.0854} & \textbf{0.0941} &  \textbf{0.1098} & \textbf{0.1137} \\
		& WMF & 0.2106 & 0.2214 & 0.2367 & 0.2401 & 0.2556 \\
		& ARMA & 0.9894 & 0.9912 & 0.9962 & 0.9995 & 1.0\\
		& MC & 1.0 & 1.0 & 1.0 & 1.0 & 1.0\\
		& Random & 0.7058 & 0.7146 & 0.7181 & 0.7126 & 0.7051 \\\hline
		\#JeSuisCharlie & hCWMF & \textbf{0.0736} & \textbf{0.0821} & \textbf{0.0876} &  \textbf{0.0912} & \textbf{0.0987} \\
		& WMF & 0.2574 & 0.2802 & 0.2946 & 0.3021 & 0.3143 \\
		& ARMA & 0.9845 & 0.9934 & 0.9971 & 1.0 & 1.0\\
		& MC & 1.0 & 1.0 & 1.0 & 1.0 & 1.0\\
		& Random & 0.7043 & 0.7051 & 0.7002 & 0.7089 & 0.7057 \\\hline
		\#OscarsSoWhite & hCWMF & \textbf{0.0834} & \textbf{0.0965} & \textbf{0.0981} & \textbf{0.1141} & \textbf{0.1185} \\
		& WMF & 0.2716 & 0.2833 & 0.2895 & 0.2934 & 0.3113 \\
		& ARMA & 0.9719 & 0.9867 & 0.9992 & 0.9998 & 1.0\\
		& MC & 1.0 & 1.0 & 1.0 & 1.0 & 1.0\\
		& Random & 0.7154 & 0.7093 & 0.7062 & 0.7091 & 0.7009 \\\hline
		\#GoodAdviceIn4Words & hCWMF & \textbf{0.0812} & \textbf{0.0853} & \textbf{0.0976} & \textbf{0.0993} & \textbf{0.1104}\\
		& WMF & 0.2401 & 0.2511 & 0.2587 & 0.2663 & 0.2877 \\
		& ARMA & 0.9924 & 0.9991 & 1.0 & 1.0 & 1.0\\
		& MC & 1.0 & 1.0 & 1.0 & 1.0 & 1.0\\
		& Random & 0.7124  & 0.7195 & 0.7071 & 0.7209 & 0.7214 \\\hline		
	\end{tabular}
\end{table*}

We have the following observations:

\begin{itemize}
	\item The performance comparison for different methods in terms of RMSE is shown in Table~\ref{tb:results1}. We observe the best results for all percentages of test set are achieved by $hCWMF$, while the next best method is the variant of our proposed method, i.e. $WMF$. The reason that two well known methods ARMA and MC fail to perform well in comparison with Random method is because the data is severely sparse while these two methods require enough data to perform well. The Random method on the other hand, regardless of the data achieves performances about \%30 all the time. 
	
	\item As stated before, for brevity, we only report the best performance of our method when the dimension of latent space $d$ is set to 10. To demonstrate the effect of $d$ on the results in Table~\ref{tb:results2}, we fix the percentage of test set to \%30 on \#MesaShooting data and vary $d$ as $\{5,10,15,20,25\}$. The observation suggests that $d = 10$ is the best dimension among others.
	
\end{itemize}

\begin{table}
	\centering
	\caption{RMSE comparison for our method on \#MesaShooting data with different dimensions of latent space and when percentage of test set is fixed to \%30}
	\label{tb:results2}
	\begin{tabular}{c|c|c|c|c}\hline
		5 & 10 & 15 & 20 & 25 \\\hline\hline
		0.1891 & \textbf{0.1181} & 0.1343 & 0.1549 & 0.1601 \\
	\end{tabular}
\end{table}

\section{Related Work}
In the realm of social network analysis~\cite{alvari2016identifying,beigi2014leveraging,alvari2013discovering,hajibagheri2012social,alvari2011detecting,hajibagheri2012community,beigi2016signed,beigi2016exploiting,beigi2016overview}, diffusion of information has been an active research area in the social network analysis~\cite{Guille:2013:IDO:2503792.2503797,IC2001,aGranovetter1978,Leskovec:2007:sdmtr,Yang:2010:MID:1933307.1934506}, however much less is known about spread of ideas in the form of hashtags. Specifically, prediction of adopting hashtags in a given time frame has been virtually an untrodden area yet. 

Most traditional works have focused on the structure and topology of the social graph. They try to maximize the spread of information and thus marketing profits by detecting the influential nodes and then leveraging viral marketing and social recommendation networks~\cite{Cha10measuringuser,Chen:2010:SIM:1835804.1835934,Goyal:2010:LIP:1718487.1718518,Kempe:2003:MSI:956750.956769,Kempe:2005:IND:2104063.2104181,Leskovec:2007:DVM:1232722.1232727}. Meanwhile some woks have modeled the temporal dynamics of information spread~\cite{ICWSM09152,Guo:2009:APU:1557019.1557064,Yang:2010:MID:1933307.1934506,Myers:2014:BDT:2566486.2568043}. In particular,~\cite{Myers:2014:BDT:2566486.2568043} studies the ways network structure reacts to users posting and sharing contents. Few have used collaborative filtering to predict the probability of a tweet to get retweeted~\cite{Zaman_predictinginformation}. Some tasks such as profiling users based on substance, style, status and social tendency has been addressed by using tweet content~\cite{conf/icwsm/RamageDL10}. 

A more recent work~\cite{Tsur:2012:WHC:2124295.2124320} proposed a hybrid linear regression based approach for predicting the spread of ideas in a given time period. In particular, they used both topology of the social graph and contents of ideas to model the propagation of ideas in Twitter by viewing hashtags as ideas and training a regression model to predict the hashtag spread in a time frame.  

On the contrary, this study does not seek modeling information propagation, rather we focus on the specific problem of predicting the adoption of hashtags in a given time period which to the best of our knowledge, has not been addressed yet.

\section{Conclusion and Future Work}
In this study, we presented an approach to address the problem of trending hashtag adoption prediction. We formulated the problem into an optimization problem and incorporated two concepts as one regularization term into the optimization equation, namely, attenuate matrix and consistency theory. We leveraged the former to model the fact that trending hashtags do not last forever, and used the latter, to investigate if one the important social theories applies to the problem of adoption of trending hashtags in Twitter. We finally used weighted low-rank matrix factorization technique to solve the problem via alternating least squares scheme which is fast for finding suboptimal matrices. Experiments demonstrate that our method outperforms other baselines for predicting trending hashtag adoption by users in future. 

Some potential avenues of future work include investigating other interesting indicator or attenuate matrices in the optimization process as well as incorporating demographics of users especially gender of the users to improve the prediction accuracy of the proposed method. In particular, we plan to replicate the study by taking into account demographics of users as potential regularization terms in the optimization equation.

\bibliographystyle{aaai}
\bibliography{ref}
\end{document}